\documentclass[prd,preprint,aps]{revtex4}

\def\D{\mathcal{D}}
\def\Tr{\mathrm{Tr}}
\def\tr{\mathrm{tr}}

\newcommand{\Slash}[1]{\!\!\not\!{#1}}
\newcommand{\SLASH}[1]{\not\!\!{#1}}

\newcommand{\Int}[1]{\int\mathrm{d^3}{#1}\,}

\begin{document}

\title{Noncommutative massive Thirring model in three-dimensional spacetime}

\author{T. Mariz, J.R. Nascimento
and R.F. Ribeiro}
\affiliation{Departamento de F\'\i
sica, Universidade Federal da Para\'\i ba,
\\
Caixa Postal 5008, 58051-970 Jo\~ao Pessoa, Para\'\i ba, Brazil}

\author{F.A. Brito}

\affiliation{Departamento de Ci\^encias F\'\i sicas e Biol\'ogicas,
\\
Universidade Regional do Cariri, 63100-000 Crato, Cear\'a, Brazil}

\date{\today}

\begin{abstract}
We evaluate the noncommutative Chern-Simons action induced
by  fermions interacting  with an Abelian gauge field in
 a noncommutative massive Thirring model in
(2+1)-dimensional spacetime. This calculation is
performed in the Dirac and Majorana representations. We
observe that in Majorana representation when $\bar\theta$ goes to
zero we do not have induced Chern-Simons term in the
dimensional regularization  scheme.
\end{abstract}

\maketitle

Noncommutative field theories have been intensively studied during
recent years --- for a review in noncommutative field theory see,
for instance, \cite{DouglasNekrasov, Szabo} and references
therein. At one hand, noncommutative spacetime, and in turn
noncommutative field theory, arise as a particular low energy
limit of string theory \cite{Connes,DouglasHull,Witten}. On the
other hand, in condensed matter theory, there is an example
leading directly to the noncommutative geometry. This is the
example of the theory of the electrons in a constant magnetic
field, projected to the lowest Landau level, which is naturally
thought of as a noncommutative field theory \cite{ja,gamboa}.
The noncommutativity geometry plays a very important role in the
context of the Hall effect \cite{prange,Susskind}.

The noncommutativity of the spacetime can be
postulated by the following commutation relation
\begin{eqnarray}
[x^{\mu},x^{\nu}] &=& i\theta^{\mu\nu}
\end{eqnarray}
with a parameter $\theta^{\mu\nu}$ which is a constant
antisymmetric tensor, and it has the canonical dimension of
inverse mass squared. As a consequence of the
noncommutativity of the spacetime coordinates one
replaces the ordinary product of functions by the Moyal product
\begin{eqnarray} \label{Mo}
f(x) \star g(x) &=&
e^{\frac{i}{2}\theta^{\mu\nu}\partial_\mu\partial^{'}_\nu}
f(x)g(x')|_{x'=x}
\end{eqnarray}
which is associative, noncommutative and satisfies the
following relation
\begin{equation}
\label{rule}
\int f(x)\star g(x) = \int g(x) \star f(x) = \int fg
\end{equation}
that is a consequence of the momentum conservation.
Thus to formulate a field theory on a noncommutative
spacetime it simply needs to replace the usual multiplication
function by the Moyal product.

In this works we are interested in analyzing the induction of the
Chern-Simons action in (2+1)-dimensional spacetime for the
noncommutative massive Thirring model.
There is a well known class of (2+1)-dimensional spacetime theories exhibiting interesting
phenomena such as exotic statistics, fractional spin and massive gauge fields \cite{red,Jackiw}.
All these phenomena are of topological nature and they can be produced when we add Chern-Simons
term to the Lagrangian which describe the system under consideration.
Another important point that has been analyzed by a great number of authors is to understand
the mechanism by which the Chern-Simons term can be generated in the usual spacetime.
Particularly, the induction of the Chern-Simons term via radiative correction has been explicitly
verified in the context of models with four-fermion (like Gross-Neveu) \cite{Rivelles} or
current-current (like Thirring) \cite{Ribeiro} self-interaction in the commutative spacetime.
The consideration of such a class of three-dimensional theories in noncommutative spacetime may reveal
new interesting physics.

We consider two different representations for the covariant
derivative of the fermions, namely, the Dirac and Majorana
representations. We discuss some relevant properties of the
induced noncommutative Chern-Simons actions
\cite{Chu,grandi_silva,Krajewski,BichlGrimstrup,ChenWu,Polychronakos,BakKim,
Sheikh,BakLee,NairPolychronakos,subir}. In the Majorana
representation we shall discuss the Chern-Simons action dependence
on the noncommutative parameter $\theta_{\mu\nu}$. Our
investigations in this paper follows the lines of the papers
\cite{AitchisonFraser1,Shifman,Fraser,AitchisonFraser2} which make
use of the derivative expansion method. In order to perform the
divergent integrals we apply the dimensional regularization.

The action for noncommutative massive Thirring model takes the form
\begin{eqnarray}\label{action}
S &=&
\int\mathrm{d}^3x\,\left[\bar\psi\star(i\Slash{\partial}-m)\psi-\frac{g}{2N}
(\bar\psi\gamma_\mu\star\psi)\star(\bar\psi\gamma^\mu\star\psi)\right].
\end{eqnarray}
The most efficient way to implement the $1/N$ expansion for this
model is to introduce an auxiliary field,
$A_\mu = (g/\sqrt{N})\bar{\psi}\gamma_\mu\star\psi$, in order to eliminate the
current-current self-interaction in Eq.(\ref{action}) so that it becomes
\begin{eqnarray}
\label{act} S[A, m] &=& \int\mathrm{d}^3x\,
\left[\bar\psi\star(i\Slash{\D}-m)\psi  + \frac{1}{2g}A_\mu\star
A^\mu  \right],
\end{eqnarray}
where the covariant derivative acting on $\psi$ is given by
\begin{eqnarray}\label{not}
\D_\mu\psi = \left\{ \begin{array}{ll}
                      \partial_{\mu}\psi + ig_{\mathrm{D}}\psi\star A_{\mu},
& \mbox{Dirac representation}, \\
                      \partial_{\mu}\psi -
ig_{\mathrm{M}}[A_{\mu},\psi]_{\star},
& \mbox{Majorana representation},
\end{array}
\right.
\end{eqnarray}
with $g_{\mathrm{D}} = \frac{1}{\sqrt{N}}$ and $g_{\mathrm{M}} = \frac{1}{2\sqrt{N}}$. The $\gamma$-matrices in (2+1) dimensions are Pauli matrices which
satisfy the well known algebra
\begin{equation}
\gamma^{\mu}\gamma^{\nu} = g^{\mu\nu} -
i\epsilon^{\mu\nu\lambda}\gamma_{\lambda}
\end{equation}
where $g^{\mu\nu}= (+, -, -)$ and $\epsilon^{123} = +1$.

The effective action $\Gamma(A, m)$ is defined as
\begin{equation}\label{effact}
Z(A,m)=e^{i\Gamma(A,m)}=\int\D\psi\D \bar\psi e^{iS[A, m]}.
\end{equation}
We shall calculate the effective action integrating on
the fermions in the Dirac representation. Let us make use
of the derivative expansion method to evaluate the one
fermion-loop effective action. To do this we substitute the
Eq.(\ref{act}) into Eq.(\ref{effact}), and use the Moyal
product, Eq.(\ref{Mo}). Thus, the effective action in the momentum
space is given by
\begin{equation}
\Gamma[A, m] =
-iN\,\Tr\,\ln{\left(\,\Slash{p}-m-g_{\mathrm{D}}\,\,e^{-\partial\times
p}\SLASH{A}(x)\right)}.
\end{equation}
The $\Tr$ stands for the trace over Dirac matrices as well as
trace over the integration in momentum and coordinate spaces.
$A_{\mu}(x)$ being dependent on the position do not commute with
functions of momentum and it is not clear how to separate out the
momentum and space dependent quantities. To do that, we shall use
the techniques of derivative expansion
\cite{AitchisonFraser1,Shifman,Fraser,AitchisonFraser2} that
proceeds as follows: $$\Gamma[A, m] = \Gamma[m] + \Gamma'[A, m]$$
where the first term is
$\Gamma[m]=-iN\,\Tr\,\ln{(\,\Slash{p}-m)}$, which does not depend
on the auxiliary field $A$. We concentrate ourselves on the second
term, which is known as Matthews-Salam determinant
\cite{MatthewsSalan}. The term is given by
\begin{equation}
 \Gamma'[A, m] =
-iN\,\Tr\,\ln{\left(1-g_{\mathrm{D}}\,\frac{1}{\Slash{p}-m}\,\,e^{-\partial\times p}\SLASH{A}(x)\right)}.
\end{equation}
The new effective action can be written as
\begin{eqnarray}\label{newaction}
\Gamma'[A, m] &=& iN\,\Tr \sum_{n=1}^\infty
\frac{1}{n}\left[g_{\mathrm{D}}\,S(p)\,\,e^{-\partial\times
p}\SLASH{A}(x)\right]^n
\end{eqnarray}
where $S(p)=(\,\Slash{p}-m)^{-1}$ is the free fermion propagator.
The exponential term that appears in the
Eq.(\ref{newaction}) is due to the noncommutativity.

The corresponding action for the quadratic term in $A_\mu$ is
given by
\begin{eqnarray}\label{actionef}
\Gamma^{(2)}[A,m] &=&
\frac{i}{2}\,\Tr\,\frac{1}{\Slash{p}-m}\,e^{-\partial\times
p}\SLASH{A}(x)\,\frac{1}{\Slash{p}-m}\,e^{-\partial'\!\times
p}\SLASH{A}(x')\Big|_{x'=x}.
\end{eqnarray}
Let us use the following  identity in order to
disentangle the $x$ and $p$ trace, i.e.,
\begin{eqnarray}\label{Exp2}
\SLASH{A}S(p)e^{-\partial'\times p}=\gamma^\mu
S(p-i\partial)e^{-\partial'\!\times (p-i\partial)}A_\mu.
\end{eqnarray}
Now we can rewrite the Eq.(\ref{actionef}) in the form
\begin{eqnarray}\label{EffAct12solved}
\Gamma^{(2)} &=& \frac{i}{2} \Int{x}
\Pi^{\mu\nu} A_\nu(x') \star A_\mu(x)\Big|_{x'=x}
\end{eqnarray}
where the tensor $\Pi^{\mu\nu}$ is given by
\begin{eqnarray}\label{pi}
\Pi^{\mu\nu} &=& \tr \int\frac{\mathrm{d}^3p}{(2\pi)^3}
S(p)\gamma^\mu S(p-i\partial)\gamma^\nu e^{-(\partial+\partial')\times p}.
\end{eqnarray}
Expanding $S(p-i\partial)$ around $p$,
\begin{eqnarray}\label{Exp3}
S(p-i\partial) &=& S(p)+S(p)i\Slash{\partial} S(p)+\cdots
\end{eqnarray}
and keeping only terms up to the first order in the
derivative $\partial$, the Eq.(\ref{pi}) is written as
\begin{eqnarray}\label{pimunu}
\Pi^{\mu\nu} &=& \tr \int\frac{\mathrm{d}^3p}{(2\pi)^3}
S(p)\gamma^\mu S(p)i\Slash{\partial}S(p)\gamma^\nu.
\end{eqnarray}
By manipulating this equation is straightforward to get
\begin{eqnarray}
\Pi^{\mu\nu} &=&
im\,\tr\int\frac{\mathrm{d}^3p}{(2\pi)^3}\frac{1}{(p^2-m^2)^3}\times\left(\,\Slash{p}\gamma^\mu\Slash{p}\Slash{\partial}\gamma^\nu+\,\Slash{p}\gamma^\mu\Slash{\partial}\Slash{p}\gamma^\nu+\gamma^\mu\Slash{p}\Slash{\partial}\Slash{p}\gamma^\nu+m^2\gamma^\mu\Slash{\partial}\gamma^\nu\! \right).
\end{eqnarray}
Now applying the $\tr$ on the $\gamma$ matrices and
solving the momentum integration, which is finite, we get
to the result
\begin{eqnarray}\label{cs1}
\Gamma^{(2)} &=& - \frac{1}{8\pi}\frac{m}{|m|}\int\mathrm{d}^3x
\epsilon^{\mu\nu\rho}\partial_\mu A_\nu \star A_\rho.
\end{eqnarray}

The corresponding action for the cubic term in $A_{\mu}$ is given by
\begin{eqnarray}\label{actioneff3}
\Gamma^{(3)}[A,m] &=&
\frac{ig_{\mathrm{D}}}{3}\,\Tr\,\frac{1}{\Slash{p}-m}\,e^{-\partial\times
p}\SLASH{A}(x)\,\frac{1}{\Slash{p}-m}\,e^{-\partial'\!\times
p}\SLASH{A}(x')\,\frac{1}{\Slash{p}-m}\,e^{-\partial''\!\times
p}\SLASH{A}(x'')\Big|_{x'=x''=x}.
\end{eqnarray}
Using the equations (\ref{Exp2}) and (\ref{Exp3}) we can
write the Eq.(\ref{actioneff3}) as
\begin{eqnarray}\label{EffAct03solved}
\Gamma^{(3)} &=& \frac{ig_{\mathrm{D}}}{3} \Int{x}
\Gamma^{\mu\rho\nu} A_\nu \star A_\rho \star A_\mu
\end{eqnarray}
where the tensor $\Gamma^{\mu\rho\nu}$ is
\begin{eqnarray}\label{gamaf}
\Gamma^{\mu\rho\nu} &=& \tr \int\frac{\mathrm{d}^3p}{(2\pi)^3}
S(p)\gamma^\mu S(p)\gamma^\rho S(p)\gamma^\nu,
\end{eqnarray}
Now applying the $\tr$ on the $\gamma$ matrices and calculating
the momentum integration in the Eq.(\ref{gamaf}) we obtain
\begin{eqnarray}\label{cs2}
\Gamma^{(3)} &=&
\frac{ig_{\mathrm{D}}}{12\pi}\frac{m}{|m|}\int\mathrm{d}^3x\epsilon^{\mu\nu\rho} A_\mu \star A_\nu \star A_\rho.
\end{eqnarray}
Finally, combining both contributions, Eq.(\ref{cs1}) and
Eq.(\ref{cs2}), we find the result
\begin{eqnarray}
\label{csT} \Gamma_{\mathrm{cs}}[A] &=&
-\frac{1}{8\pi}\frac{m}{|m|}\int\mathrm{d}^3x
\epsilon^{\mu\nu\rho}\left(\partial_\mu A_\nu \star A_\rho -
\frac{2ig_{\mathrm{D}}}{3}A_\mu \star A_\nu \star A_\rho \right),
\end{eqnarray}
which is the Chern-Simons action for the Dirac representation.
Observe that even in the abelian case we have a cubic term,
similar to the Chern-Simons action for ordinary non-abelian case.

In the Majorana representation the effective action takes
the form
\begin{equation}\label{novaacaoeffmaj}
\tilde{\Gamma}'[A,m] = i\frac{N}{2}\Tr\sum_{n=1}^\infty
\frac{1}{n}\left[g_{\mathrm{M}}
\,S(p)\,\left(e^{-\partial\times p}
-e^{\partial\times p}\right)\SLASH{A}(x)\right]^n.
\end{equation}
The corresponding action for the quadratic term in $A$ is given by
\begin{eqnarray}\label{novaacaoeff21maj}
&\tilde{\Gamma}^{(2)} &= \frac{i}{4}\frac{g_{\mathrm{M}}^2}{g_{\mathrm{D}}^2} \int\mathrm{d}^3x\,\tr
\int\frac{\mathrm{d}^3p}{(2\pi)^3}
S(p)\gamma^\mu S(p-i\partial)\gamma^\nu \nonumber\\
                        &\times&\left[\left(e^{(\partial+\partial')\times p}-
e^{-(\partial-\partial')\times p}\right) A_\mu(x) \star A_\nu(x') +
\left(e^{-(\partial+\partial')\times p}-
e^{(\partial-\partial')\times p}\right) A_\nu(x') \star
A_\mu(x)\right]_{x'=x}.
\end{eqnarray}
Rewriting the Eq.(\ref{novaacaoeff21maj}) in the momentum
space we find
\begin{eqnarray}\label{novaacaoeff22maj}
\tilde{\Gamma}^{(2)} &=&
\frac{i}{2}\frac{g_{\mathrm{M}}^2}{g_{\mathrm{D}}^2}\int\frac{\mathrm{d}^3k}{(2\pi)^3}\left(\tilde{\Pi}^{\mu\nu}_{\mathrm{p}}+
\tilde{\Pi}^{\mu\nu}_{\mathrm{np}}\right) A_\mu(k) A_\nu(-k),
\end{eqnarray}
where the contributions from planar graphics are given
by $\tilde{\Pi}^{\mu\nu}_{\mathrm{p}}=\Pi^{\mu\nu}$, and the
contributions from non-planar graphics are
written as
\begin{equation}\label{pinp}
\tilde{\Pi}^{\mu\nu}_{\mathrm{np}} =
-\frac12\,\tr\int\frac{\mathrm{d}^3p}{(2\pi)^3}
\left(S(p)\gamma^\mu S(p)\gamma^\nu+S(p)\gamma^\mu S(p)\Slash{k}S(p)\gamma^\nu +
 \cdots\right) \left(e^{2ik\times p}+e^{-2ik\times p}\right),
\end{equation}
where we have used the Eq.(\ref{Exp3}). Thus, keeping only linear term in
the derivative, the Eq.(\ref{pinp}) turns out to be
\begin{eqnarray}\label{pinp1}
\tilde{\Pi}^{\mu\nu}_{\mathrm{np}} &=&
-m\,\tr\int\frac{\mathrm{d}^3p}{(2\pi)^3}\frac{e^{i\bar{\theta}.p}}{(p^2-m^2)^3}
\times\left(\,\Slash{p}\gamma^\mu\Slash{p}\Slash{k}\gamma^\nu+\,\Slash{p}\gamma^\mu\Slash{k}\Slash{p}
\gamma^\nu+\gamma^\mu\Slash{p}\Slash{k}\Slash{p}\gamma^\nu+m^2\gamma^\mu\Slash{k}\gamma^\nu\! \right),
\end{eqnarray}
with $\bar{\theta}^\alpha=\theta^{\alpha\beta}k_\beta$. Making use of the identities $\gamma^\alpha\gamma_\alpha=3$,
$$\tr(\gamma^\mu\gamma^\nu\gamma^\alpha\gamma^\beta\gamma^\rho)=
g^{\mu\nu}\tr(\gamma^\alpha\gamma^\beta
\gamma^\rho)+
g^{\alpha\beta}\tr(\gamma^\mu\gamma^\nu\gamma^\rho)+
g^{\mu\rho}\tr(\gamma^\alpha\gamma^\beta\gamma^\nu)-
g^{\nu\rho}\tr(\gamma^\alpha\gamma^\beta\gamma^\mu)$$
and of the relations for the integrals \cite{Brandt}
\begin{eqnarray}
\int\frac{\mathrm{d}^3p}{(2\pi)^3}\frac{e^{i\bar{\theta}.p}}{(p^2-m^2)^3}&=&-
\frac{i}{32\pi}\frac{1}{m^2|m|}\left(1+m|\bar{\theta}|\right)e^{-m|\bar{\theta}|} \nonumber\\
\int\frac{\mathrm{d}^3p}{(2\pi)^3}\frac{p_\alpha p_\beta\,
e^{i\bar{\theta}.p}}{(p^2-m^2)^3}&=&
\frac{i}{32\pi}\frac{1}{|m|}\left(g_{\alpha\beta}-m|\bar{\theta}|\frac{\bar{\theta}_\alpha
\bar{\theta}_\beta}
{\bar{\theta}^2} \right)e^{-m|\bar{\theta}|},
\end{eqnarray}
we can express the Eq.(\ref{pinp1}) in the form
\begin{eqnarray}\label{resultadopiadj}
\tilde{\Pi}^{\mu\nu}_{\mathrm{np}} &=&
-\frac{k_\rho}{4\pi}\frac{m}{|m|}e^{-m|\bar{\theta}|}\epsilon^{\mu\nu\rho} -
\frac{k_\rho}{16\pi}\frac{m^2}{|m|}|\bar{\theta}|e^{-m|\bar{\theta}|}\left[\epsilon^{\mu\nu\rho}-
\frac{\bar{\theta}_\alpha}{\bar{\theta}^2}\left(\bar{\theta}^\mu\epsilon^{\alpha\nu\rho}+
\bar{\theta}^\nu\epsilon^{\mu\alpha\rho}+ \bar{\theta}^\rho\epsilon^{\mu\nu\alpha} \right)\right].
\end{eqnarray}
The second term of this equation disappears when we use the relation
\begin{equation}
\bar{\theta}_\alpha\left(\bar{\theta}^\mu\epsilon^{\alpha\nu\rho}+
\bar{\theta}^\nu\epsilon^{\mu\alpha\rho} + \bar{\theta}^\rho\epsilon^{\mu\nu\alpha}\right)=\bar{\theta}^2
\epsilon^{\mu\nu\rho},
\end{equation}
so that the non-planar graphics contribute as
\begin{equation}
\tilde{\Pi}^{\mu\nu}_{\mathrm{np}} =
-\frac{k_\rho}{4\pi}\frac{m}{|m|}e^{-m|\bar{\theta}|}\epsilon^{\mu\nu\rho}.
\end{equation}
Now substituting the expressions obtained for planar and
non-planar graphics into Eq.(\ref{novaacaoeff21maj}), we
obtain the term
\begin{eqnarray}
\tilde{\Gamma}^{(2)} &=& -\frac{1}{8\pi}\frac{m}{|m|}\frac{g_{\mathrm{M}}^2}{g_{\mathrm{D}}^2}\int\mathrm{d}^3x \left(1-e^{-m|\bar{\theta}|}\right)\epsilon^{\mu\nu\rho}\partial_\mu A_\nu \star A_\rho.
\end{eqnarray}

The corresponding action for the cubic term, in the momentum
space, is given by
\begin{equation}\label{novaacaoeff31maj}
\tilde{\Gamma}^{(3)} = \frac{i}{3}\frac{g_{\mathrm{M}}^3}{g_{\mathrm{D}}^2}
\int\frac{\mathrm{d}^3k}{(2\pi)^3}\int\frac{\mathrm{d}^3k'}{(2\pi)^3}
\left(\tilde{\Gamma}^{\mu\rho\nu}_{\mathrm{p}} +
\tilde{\Gamma}^{\mu\rho\nu}_{\mathrm{np}} \right) e^{-ik'\times
k}A_\mu(k')A_\rho(k)A_\nu(-k-k').
\end{equation}
The planar diagrams contribute as
$\tilde{\Gamma}^{\mu\rho\nu}_{\mathrm{p}}=\Gamma^{\mu\rho\nu}$
and the expression for non-planar diagrams is given by
\begin{equation}\label{gammaadj}
\tilde{\Gamma}^{\mu\rho\nu}_{\mathrm{np}} = -\frac12\,\tr \int\frac{\mathrm{d}^3p}{(2\pi)^3} S(p)\gamma^\mu
 S(p)\gamma^\rho S(p)\gamma^\nu \left(e^{ik''\times p}+e^{-ik''\times p}\right),
\end{equation}
where $k''=-k-k'$. After manipulating this equation we can see that it is very similar to the Eq.(\ref{pinp1})
and therefore the Eq.(\ref{novaacaoeff31maj}) takes the form
\begin{equation}
\tilde{\Gamma}^{(3)} =
\frac{ig_{\mathrm{M}}}{12\pi}\frac{m}{|m|}\frac{g_{\mathrm{M}}^2}{g_{\mathrm{D}}^2}\int\mathrm{d}^3x
\left(1-e^{-m|\bar{\theta}|}\right)\epsilon^{\mu\nu\rho} A_\mu \star A_\nu \star A_\rho.
\end{equation}
Finally, we combine the planar and non-planar contributions
to get the Chern-Simons term for the Majorana representation,
i.e.,
\begin{equation}\label{S}
\tilde{\Gamma}_{\mathrm{cs}} = -\frac{1}{8\pi}\frac{m}{|m|}\frac{g_{\mathrm{M}}^2}{g_{\mathrm{D}}^2}\int\mathrm{d}^3x \left(1-e^{-m|\bar{\theta}|}\right)\epsilon^{\mu\nu\rho}\left(\partial_\rho A_\mu \star A_\nu-\frac{2ig_{\mathrm{M}}}{3}A_\mu \star A_\nu \star A_\rho \right).
\end{equation}
In this equation we observe that the Chern-Simons action
$\tilde{\Gamma}_{\mathrm{cs}}$ is vanish  when we take $\bar\theta \to 0$.
This means that UV/IR mixing, which is a characteristic of some
noncommutative field theories, is not observed. Also, as we can
see in the Eq.(\ref{S}) if $m \to 0 $, massless fermions, we do
not have induced Chern-Simons action. This happens due to fact
that the term $m\bar\psi \psi$ is vanish and therefore the parity
symmetry is not broken. As in the commutative case parity
violation is prerequisite to generate Chern-Simons term. We can
still consider $m|\bar\theta|\to$ fixed, such that for $m\ll1$ one
has $|\bar\theta|\gg1$. In this limit the Eq.(\ref{S}) does not
vanish and we predict light fermions traveling in a strongly
noncommutative spacetime. Alternatively, from the dispersion
relation, $E^2\!=\!c^2p^2+c^4m^2+f(p^\mu\bar\theta_{\mu})$, where
$\bar\theta_{\mu}\!=\!\theta_{\mu\nu}p^\nu$, $p_\mu$ here is the
external momentum and $f$ depends on the spin and charges of the
particle
--- see, for instance, \cite{amelino} and references therein ---
we see that the effects of the tensor $\theta_{\mu\nu}$ for light
fermions, i.e., $m\ll1$, with the same spin and charges, are
stronger than the effects for heavy fermions. The effects of
$\theta_{\mu\nu}$ on such particles could be, for instance,
similar to the ones that a birefringent crystal induces in a light
beam \cite{amelino}.

We summarize our work recalling that we have investigated the
induction of a Chern-Simons action by the massive Thirring model
in noncommutative spacetime, in the Dirac and Majorana
representations. In the Dirac representation the Chern-Simons
action is similar to ordinary non-abelian case.

As expected, the induced Chern-Simons action is gauge invariant
under large gauge transformations. This happens due to the fact
that we have considered parity violation that was originated in
the fermion mass term. Another point that we have observed is that
in the Majorana representation when we take $\bar\theta \to 0$ the
Chern-Simons action is vanish and therefore UV/IR mixing does not
appear.

Finally, we observe according to Eq.(\ref{S}), by fixing
$m|\bar\theta|$, we can predict that light fermions are more
sensitive to the spacetime noncommutativity than heavy fermions.

\acknowledgments

We would like to thank D. Bazeia and V. O. Rivelles  for useful
discussions and CNPq, PROCAD for partial support.

\end{document}